\documentclass[12pt]{iopart}
\usepackage{iopams}
\usepackage{epsfig}

\begin{document}

\title[An hybrid Tykhonov method for neutron spectrum unfolding]
{An hybrid Tykhonov method for neutron spectrum unfolding}

\author{Olivier Besida\dag\
\footnote[3]{To
whom correspondence should be addressed (olivier.besida@hep.saclay.cea.fr)}
}

\address{\dag\ CEA/DAPNIA/SENAC,
CE Saclay, 91191 Gif sur Yvette, France}

\begin{abstract}
An hybrid iterative Tykhonov regularization approach with an accelerating algorithm is considered. This method is illustrated by two
neutron spectrum unfoldings measured with a Bonner Sphere system.
\end{abstract}


\submitto{\IP}

\maketitle

\section{Introduction}
Following the ideas of A.N. Tykhonov about ill-posed problems inversion, we extend the iterative regularization method by
including alternatively between two iterations least square gradient looking terms, and with a modification of the parametrization of the regularizing term.
The merits of this approach are a fast convergence to a satisfying solution, stable and unsensitive to the noise fluctuations,
and a complete independence with respect to the initial guess.
With this method, we unfold one dimensional data set of measurements with a Bonner
Sphere system to obtain a wide neutron spectrum estimation. Two experimental cases are presented : a complex thermal neutron wide spectrum and a single monoernegetic 565 keV neutron spectrum.
The evolution of the criteria parameter is discussed and also the systematics and statistics error bars calculations.

\section{Theoretical approach}
Once we have a set of 11 measurements, by spline interpolation, we obtain a vector $y$ of dimension 64 $(y \in \mathbb{R}^{+64})$;
this is justified by the smoothness of the data set and by the convolution that erodes most of the details of the initial
 spectrum. The measurement function $y \in C^\infty $ and is very smooth. The spline interpolation produces a
64 bin vector, with continous derivatives at each of the 11 knots and local second degree polynomial interpolation.
These 11 measurements corresponds to the responses to a neutron spectrum of an Helium$^3$ neutron detector with
various polyethylene spheres, that are equivalent to different filters with response matrix A. With $x \in \mathbb{R}^{+64}$
as the unknown neutron spectrum, $ y $ as the measurements, the convolution equation of the system stands as :
\begin{eqnarray}
 y=Ax\\
 x \in \mathbb{R}^{+64}\qquad y \in \mathbb{R}^{+64} \qquad A
\in \mathbb{R}^{+64}\times\mathbb{R}^{+64}  \nonumber
\end{eqnarray}
As the problem is ill-posed, neither the direct inversion, nor the least square solution are satisfying solutions, because
they are irregular and highly sensitive to the fluctuations of the measurement's uncertainties. That means that
continuous variations of $y$ does not lead to continuous variations in $x$.
\begin{eqnarray}
\mbox{ \em Direct inversion : } \nonumber\\
 \qquad \mbox{ \em estimator :  }  \tilde x  = A^{-1} y    \label{E:eqn2}\\
 \mbox{ \em Least square solution : }\nonumber \\
 \qquad \mbox{\em criterion :  }  Q =|Ax-y|^2  \label{E:eqn3}\\
 \qquad \mbox{\em estimator :  }  \tilde x=[A^t A]^{-1} A^t y  \label{E:eqn4}\\
 \qquad ( A^t  \mbox{\em  : Transposed matrix} ) \nonumber
\end{eqnarray}

\subsection{Iterative Tykhonov approach}
A better solution can be achieved through the scheme of the basic Tykhonov
method \cite{Tykhonov1, Tykhonov2, Tykhonov3, Tykhonov4,Riley,Anykeyev} ,
 which is more regular and less sensitive to noisy fluctuations.
A regularization term is added to the least square criterion through a Lagrange multiplicator method.
This term ensures the continuity of the solution and its closedness to an initial guess $x_0$, which can be
a "not so bad" first solution.
\begin{eqnarray}
& Q_\lambda =|Ax-y|^2+\lambda |x-x_0|^2  \qquad 	\lambda \in [0,+\infty]     \label{E:eqn5}\\
& \tilde x=[A^t A+\lambda I]^{-1}[A^ty+\lambda x_0]    \label{E:eqn6}
\end{eqnarray}
The demonstration of the calculation of the estimator (equation 6) can be found as follows \cite{Bertero} : \\
Assume $\tilde x$ to be the global minimum of the criterion $Q_\lambda$.
\begin{eqnarray}
& \forall x \in \mathbb{R}^{+64} : Q_{\lambda,\tilde x} \leq  Q_{\lambda,x} \label{E:eqn7}
\end{eqnarray}
This is true, moreover in the case of small variations around $\tilde x$ :
\begin{eqnarray}
& x=\tilde x+\alpha h \qquad h \in  \mathbb{R}^{+64} \qquad \alpha \in \mathbb{R} \nonumber \\
& Q_{\lambda,\tilde x} \leq Q_{\lambda,\alpha,h}  \nonumber \\
& \Leftrightarrow Q_{\lambda,\alpha,h} - Q_{\lambda,\tilde x} \geq 0  \nonumber \\
& \Leftrightarrow |A(\tilde x +\alpha h)-y|^2 +\lambda|\tilde x +\alpha h-x_0|^2-|A\tilde x-y|^2 -
\lambda|\tilde x -x_0|^2 \geq 0  \nonumber \\
& \mbox{\em Due to the definition of the scalar product :} \nonumber \\
& \Leftrightarrow \alpha^*(Ah)^t [A\tilde x -y]+\alpha[A\tilde x -y]^t Ah+|\alpha|^2|Ah|^2+ \alpha^*\lambda h^t [\tilde x -x_0 ]+ \nonumber \\
& \alpha\lambda[\tilde x-x_0]^th+\lambda|\alpha|^2|h|^2 \geq 0  \nonumber \\
& \Rightarrow (\alpha^* h^t )(A^t[A\tilde x -y]+\lambda[\tilde x -x_0 ])+\alpha([A\tilde x-y]^tA+\lambda[\tilde x-x_0]^t)h \geq 0  \nonumber \\
& \qquad \mbox{\em That is independent of } h^t  \mbox{ and } \alpha \mbox{ ; then it implies : }  \nonumber \\
& \Rightarrow (A^tA\tilde x-A^ty+\lambda \tilde x-\lambda x_0)=0 \nonumber  \\
& \Leftrightarrow [A^tA+\lambda I]\tilde x-[A^ty+\lambda x_0]=0 \nonumber \\
& \Leftrightarrow \tilde x=[A^tA+\lambda I]^{-1}[A^ty+\lambda x_0]  \nonumber
\end{eqnarray}
The uniqueness and existence of a bounded solution is also proved in Bertero \cite{Bertero,Neumaier}.
The Tykhonov iterative approach is achieved by successive refinement of the solution with several iterations, beginning with an
initial guess $ x_0 $, the solution $x_1$ satisfies better the criterion of Tykhonov than $x_0$, and so on :
\begin{eqnarray}
& Q_{\lambda ,n} =|Ax_{n}-y|^2+\lambda |x_{n}-x_{n-1}|^2  \qquad 	\lambda \in [0,+\infty]     \label{E:eqn8}\\
& \tilde x_n=[A^t A+\lambda I]^{-1}[A^ty+\lambda x_{n-1}]    \label{E:eqn9} \\
&  Q_{\lambda ,1} \geq Q_{\lambda ,2}  \geq ... \geq Q_{\lambda ,n}  \geq Q_{\lambda ,n+1}  \geq ... \nonumber
\end{eqnarray}

\subsection{Modification of the running parameter $\lambda$}
In order to help the computer calculation, a slight modification is done with the modification of the running
interval of the parameter $ \lambda $. We also decided that the two terms of the $Q_{\lambda,n} $ criterion have to be of equal importance :
\begin{eqnarray}
& \lambda \in [0,+\infty] \longmapsto \lambda \in [0,1]  \label{E:eqn10}\\
& Q_{\lambda ,n}=(1-\lambda)|Ax_n-y|^2+\lambda |x_n-x_{n-1}|^2  \qquad   \lambda \in [0,1]   \label{E:eqn11}\\
& \tilde x_n=[(1-\lambda)A^t A+\lambda I]^{-1}[(1-\lambda)A^t y+\lambda x_{n-1}]   \label{E:eqn12}
\end{eqnarray}
Let's have $ x_0 $ as an initial guess, and lets compute $ x_1 $.
The case $ \lambda =0 $ gives :
\begin{eqnarray}
& Q_{0,1}=|Ax_1-y|^2     \label{E:eqn13}\\
& \Rightarrow \tilde x_{0,1}=[A^t A]^{-1}A^t y   \qquad  \mbox{ \em   (least square solution)}  \label{E:eqn14} \\
& \Rightarrow  0 \leq Q_{0,1}           \label{E:eqn15}
\end{eqnarray}
On the other side, the case $ \lambda =1 $ gives	:
$ Q_{1,1}=|x_1-x_0|^2 $	with the solution $ x_1=x_0 $ then $ Q_{1,1}=0 $ then $ \lambda =1 $ is a minimorum and
all the iterations stand for an identity operator, then $\forall n \in \mathbb{N}^* \qquad  x_n=x_0 $ independently of $ x_0 $ ! So this method gives always a wrong
solution, because all initial guesses are available solutions without any restriction. We fixed this problem by modifying the criteria in the following way :
\begin{eqnarray}
&Q_{\lambda ,n}={(1-\lambda)|Ax_n-y|^2+\lambda |x_n-x_{n-1}|^2 \over (1-\lambda)\lambda} \qquad  \lambda \in ]0,1[ \label{E:eqn16}\\
&\Leftrightarrow	Q_{\lambda ,n}={|Ax_n-y|^2 \over \lambda}+{|x_n-x_{n-1}|^2 \over {1-\lambda}}  \label{E:eqn17}
\end{eqnarray}
the estimator is unchanged :
\begin{equation}
\tilde x_n=[(1-\lambda)A^t A+\lambda I]^{-1}[(1-\lambda)A^t y+\lambda x_{n-1}] \label{E:eqn18}
\end{equation}
but $ Q_{0,n}=+ \infty $ and $ Q_{1,n}=+\infty $\\
Also we can note : $ \exists  \lambda \in ]0,1[ $ , $ Q_{\lambda,n}<+\infty $
$ \Leftrightarrow \exists $ an actual minimorum of $ Q_{\lambda ,n} $ and the false minimorum $ Q_{1,n} $ disappears!

\subsection{Possible generalization to derivatives of order m}
Basically, the Tykhonov method can be generalized to derivatives of order m of $x_n$ instead of using the simple
 squared modulus \cite{Giovanelli}.
These derivatives $D^{(m)}x_n$ are the derivatives of order m of the  spectrum $x_n(E)$ regarding the energy binning $E$ :
$D^{(m)}x_n= {d^{(m)}\over {dE}}x_n(E)  $ \\
This generalization to greater orders of derivatives is given by :
\begin{eqnarray}
& Q_{\lambda ,n,m}={|Ax_n-y|^2 \over \lambda }+{|D^{(m)}x_n-D^{(m)}x_{n-1}|^2 \over 1-\lambda} \label{E:eqn19} \\
& \tilde x_n=[(1-\lambda)A^t A+\lambda D^{(m)t}D^{(m)}]^{-1}[(1-\lambda)A^t y+\lambda D^{(m)t}D^{(m)}x_{n-1}]  \label{E:eqn20}
\end{eqnarray}
But we have found that the convergence to a stable solution is rather better and faster with a simple function without the $m^{th}$ derivative, so
we decided to restrict our calculation only to order 0.

\subsection{Hybridation with the least square method}
Three main concerns guided the development of this unfolding iterative method : a fast converging method, a better solution than the classical
Tykhonov method or than the least square method and the last one, a complete independence of the solution regards to the initial guess.
The iterative Tykhonov approach ensures the last concern but not the first two. So we decided to introduce
some accelerating (focusing) algorithm between two classical Tykhonov iterations.
Let suppose that $x_{n-1}$ and $x_n$ are two successive solutions obtained with this Tykhonov algorithm, we create some
kind of derivative term regarding the iteration parameter n:
\begin{eqnarray}
& \partial x_n/ \partial n=x_n-x_{n-1}  \nonumber \\
& \lim _{n \to +\infty} { \partial x_n \over { \partial n } }= 0  \Leftrightarrow \exists x \in \mathbb{R}^{+64} ; \lim_{ n \to +\infty}x_n=x \nonumber \\
& \mbox {Then we minimize } Q_{\mu,n+1} \nonumber \\
& Q_{\mu,n+1}=|A x_{n+1} -y |^2 \mbox{ with } x_{n+1}=x_n+\mu {\partial x_n \over { \partial n }}; \mu \in \mathbb{R} \nonumber \\
& \Leftrightarrow Q_{\mu,n+1}=|A (x_n+\mu {\partial x_n \over {\partial n}} ) -y |^2 \label{E:eqn21}
\end{eqnarray}
So after every two Tykhonov unfoldings, we compute one solution closed to the least square solution.

At the end of the convergence, the reconstructed spectrum $y_n=Ax_n$ is closer to
the measured spectrum $y$ than with use of the classical Tykhonov iterative scheme,
which presents the compromise of an enhancement of the stability of the solution toward
the noise fluctuations while the price to pay is a loss of the accuracy \cite {OFTA}  and an excursion from the measured spectrum $y$.
By the way, this approach converges faster than a classical iterative Tykhonov method.
This hybridation algorithm acts as a focalisation method with respect to the requested measurement $y$, and also as a great accelerator of convergence.

\subsection{Iteration schema}
The iterative schema can be described in the following way. Let's have $x_0$ as any initial guess.
\begin{eqnarray}
 \mbox{ \em iteration 1 :  }&  \nonumber \\
&Q_{\lambda ,1}={|Ax_1 -y|^2 \over \lambda }+{ |x_1 -x_0 |^2 \over (1-\lambda)}  \label{E:eqn22}\\
&\tilde x_1=[(1-\lambda)A^t A+\lambda I]^{-1}[(1-\lambda)A^t y+\lambda x_0]    \label{E:eqn23}\\
\mbox{ \em iteration 2 :  }&  \nonumber \\
& Q_{\lambda,2}={|Ax_2-y|^2 \over \lambda}+{|x_2- x_1|^2 \over (1-\lambda) }    \label{E:eqn24}\\
& \tilde x_2=[(1-\lambda)A^t A+\lambda I]^{-1}[(1-\lambda)A^t y+\lambda  x_1]    \label{E:eqn25}\\
\mbox{ \em iteration 3:   } & \nonumber \\
&  Q_{\mu,3}=|A (x_2+\mu {\partial x_2 \over { \partial n }}) -y |^2 \label{E:eqn26} \\
& \tilde x_3=x_2+\mu {\partial x_2 \over { \partial n }} \label{E:eqn27} \\
 \mbox{ \em iteration 4 :  } &  \nonumber \\
& Q_{\lambda ,4}={|Ax_4-y|^2 \over \lambda}+{|x_4-x_3|^2 \over 1-\lambda }     \label{E:eqn28}\\
& \tilde x_4=[(1-\lambda)A^t A+\lambda I]^{-1}[(1-\lambda)A^t y+\lambda x_3]   \label{E:eqn29}\\
&\vdots \nonumber \\
 \mbox{ \em iteration n :  } &    n\equiv 1 \mbox{ modulo } 3  \nonumber \\
& Q_{\lambda ,n}={|Ax_n-y|^2 \over \lambda}+{|x_n-x_{n-1}|^2 \over 1-\lambda}     \label{E:eqn30}\\
& \tilde x_n =[(1-\lambda)A^t A+\lambda I]^{-1}[(1-\lambda)A^t y+\lambda x_{n-1}]  \label{E:eqn31} \\
 \mbox{ \em iteration n+1 :  } &    n+1\equiv 2 \mbox{ modulo } 3  \nonumber \\
& Q_{\lambda ,n+1}={|Ax_{n+1}-y|^2 \over \lambda}+{|x_{n+1}-x_{n}|^2 \over 1-\lambda}     \label{E:eqn32}\\
& \tilde x_{n+1} =[(1-\lambda)A^t A+\lambda I]^{-1}[(1-\lambda)A^t y+\lambda x_{n}]  \label{E:eqn33} \\
 \mbox{ \em iteration n+2 :  } &    n+2\equiv 0 \mbox{ modulo } 3  \nonumber \\
&  Q_{\mu,n+2}=|A (x_{n+1}+\mu { \partial x_{n+1} \over { \partial n }} ) -y |^2 \label{E:eqn34} \\
& \tilde x_{n+2}=x_{n+1}+\mu { \partial x_{n+1} \over \partial n } \label{E:eqn35} \\
&\vdots \nonumber
\end{eqnarray}

We have seen with the help of a computer that in all cases, this iterative schema converges to
one unique solution, but we have not build the mathematical proof of this convergency!
Practically, on a DEC-Compaq Alpha Workstation working with double precision 64 bits numbers,
the resulting estimators are always stable after 300 iterations  ( 1 minute of computing time)
and oftenly around 80 iterations, it is already stable. That is why we decided to put a cut on
the maximum number of iterations at 300. We also decided to restrict the binning
(in energy and interpolated sphere) over 64 bins, because working with 32 or 128 bins gives either a
too bad energy resolution or a too slow computation, thus 64 seems to be an optimum.

\subsection{Positivity of the solution}
	Another important aspect of the unfolding is treated without any subtility: positivity \cite{Giovanelli}.
Of course each bin of the spectrum must have a positive content, on the other case the spectrum would be unphysical.
The positivity is introduced abruptly at each step of the iteration with the following non-linearity:
\begin{equation}
\tilde x_{n,i} \longmapsto Sup[0,\tilde x_{n,i}] \label{E:eqn36}
\end{equation}

\subsection{Calculation of the statistics and systematics fluctuations}
	The statistics fluctuations are estimated with a Monte-Carlo method. For one set of 11 measurements with the
 Bonner Sphere system,11 uncertainties of these measurements are calculated taking into account the number of
observed counts and the duration of the measures. The propagation through the unfolding process of the statistics
fluctuations of the measurements is computed with 100 sets of 64 data $y$, which are interpolated with the spline
method from these 11 knots. The random generation of the 11 knots is performed with a gaussian law centered onto the
11 actual measurements and with a standard deviation corresponding to the error bars directly measured with the
instrument. The unfolding of the 100 data sets $y$ gives 100 $\tilde x $ estimators of the neutron spectrum.
The fluctuations of the 100 spectra $\tilde x$ correspond to the propagation of the statistics fluctuations through
the unfolding process.
\begin{equation}
\sigma_{j,statistics}^2= { { \sum^{N}_{i=1} |\tilde x_{i,j}^2 - <\tilde x_{i,j}> ^2|} \over {N} } \qquad N=100 \label{E:eqn37}
\end{equation}
Also, to the 100 $ \tilde x$ spectra correspond 100 reconstructed measurements $\tilde y= A \tilde x $.
The reconstructed measurements $\tilde y $ are reinjected into the unfolding process giving new estimators of the spectra $\tilde x_{new}$ and twice
reconstructed measurements $\tilde y_{new}$. The average of the absolute difference of the old and of the new
spectra gives the systematics errors of the unfolding spectra :
\begin{equation}
\sigma_{j,systematics}^2= { { \sum^{N}_{i=1} |\tilde x_{i,j} - \tilde x_{i,j,new} |^2} \over {N} } \qquad N=100 \label{E:eqn38}
\end{equation}
These two calculations of uncertainties are performed with 100 samples of Monte-Carlo generated
data sets; this figure seems low due to the signal to
 noise ratio of 10 \%, nevertheless the precision is sufficient for our purpose and moreover
it is highly time consuming and we cannot afford greater computation. Also some tricks are used to
 avoid spending too much time recalculating 100 times almost
the same things. Of course the $\lambda$ and $\mu$ parameters minimizations are very
expensive on the CPU time balance, so the evolution
with the iterations of the $\lambda _n$ (for example) is memorized during the first pass, while for the next
passes the minimization of the $\lambda$ parameter  is performed at the
$n^{th}$ iteration in the neighboorhood of the $\lambda_n$ first evaluation. Another trick is used, oftenly after
50 or 80 iterations the $\tilde x_n$ and $\tilde y_n$ figures are stables
due to the precision of the computation, so each successive iteration appears as an Identity operator, then  at this
moment the iterative process is stopped.

\subsection{The initial guess}
This kind of unfolding method shows several advantages. The most important one is the independance of the final
solution regarding the initial guess.
Thus any distribution can be used as an initial guess. We have tested several different initial guesses : random numbers,
flat functions $x_0=1$ and the intercorrelation product between the response $y$ and the discrete response
functions of matrix $A$.
\begin{equation}
x_{0,j}=<y_i,A_{i,j}\delta_j> \qquad \mbox{ \em $\delta_j$ : dirac function } \label{E:eqn39}
\end{equation}
The strong point of this method is that using all these different initial guesses the iterative approach converges
to exactly the same result.
For a reason of fast and easy computation, we decided to use a flat initial guess : $x_0=1$.

\subsection{Remarks about the minimization algorithms}
	To circonvene the problem of the minimization with $\lambda$ of the $Q(\lambda)$ criteria, we used different technics
in order to maintain a high precision on the real $\lambda$ and a reasonable time of calculation for each iteration.
The minimization of the criteria is performed in different ways to optimize this calculation time. Firstly, a multiple step scanning and
secondly, a dichotomy algorithm are performed. The first solution of the first set $y$ is calculated carefully, while the other
solutions computed to estimate the systematic and the statistic uncertainties, does begin their minimization at each step,
with initial figures in the neighborhood of the first $\lambda_n$ sequence calculated previously. Of course, at this point the focusing on the right
$\lambda$ is performed with scanning and dichotomy algorithms but with less steps, that avoid to spoil too much computation time.

\section{Experimental approach}
To illustrate this unfolding technics, we presents two examples of unfolding: the first one is a thermal neutron spectrum,
 the second one is a 565 keV  mono-energetic neutron spectrum.

\subsection{Unfolding a thermal neutron spectrum from the Sigma installation}
	This thermal neutron spectrum is produced near the Sigma facility at (IRSN-Cadarache), it consists of an assembling
of several AmBe neutron sources surrounded by some graphite shielding. The expected spectrum is an heavy thermal component
with some residual peaks between 1 and 3 MeV. Figures 1.a and 1.b shows precisely what was expected : a huge thermal neutron part
and a small high energy component that goes from 100 keV up to 10 MeV. If the lower limit is physically explainable due
to the tail of the decelarated neutrons from the high energy peaks, the upper bound around 10 MeV is not realistic and it is just
due to the intrinsic resolution of the Bonner Sphere method itself. While figure 1.a shows the statistics error bars, figure 1.b presents
the systematics unfolding error bars; in both case they are extremely tiny : the global statistics uncertainties is  0.11\%  while
the global systematics errors of the unfolding is 3.6\%. These very small uncertainties are mainly due to the very high number of counts
registered for most of the sphere, this can be seen on figure 1.c, which represents the measured counts on the various sphere,
(plain line), while the dot line corresponds to the recalculated spectrum after unfolding. These two shapes are in perfect agreement together.
The last figures, shows one of the very specific parameter of the evolution of the unfolding process : $1-\lambda $
 {\em vs.} $ iteration $  $ number $.
This shows that after only 20 iterations this parameter is already very low $(10^{-15})$, that means that the iterative operator is very closed to the Identity operator,
and the current solution is very closed to the final solution. This shape is very easy to unfold, this is why after only 78 iterations, the unfolding
process is over.
The main aspect of this unfolding is its ability to reveal some realistic details of the shape here a double structured shape with confortable uncertainties.

\subsection{Unfolding a monoenergetic neutron spectrum at 565 keV}

The Bonner Sphere system was operated in the beam of a Van de Graaf accelerator producing 565 keV mono-energetic
neutrons at CEA (Bruy\`eres le Ch\^atel). The expected spectrum is a single energy peak enlarged by the energy resolution of the device eventually
 surrounded by some low energy diffused neutrons. The background of the low energy diffused neutrons is substracted from the measurements
thanks to other measures undertook with copper-polyethylene shadow cones between the target that produces the
neutrons and the detector. Figure 2c shows a light discrepancy between the measurements and the reconstructed data
for the values between the 4 and the 7 inches spheres; this is probably due to the various efficiencies of the
shadowing cones for small and big spheres. Anyway the global discrepancy between these two curves is 1.2 $\%$.
Anywhere else, for the right and left wings, the reconstructed distribution is in good agreement with the original data.
The unfolded spectrum (fig 2a and 2b), clearly shows a mono-energetic spectrum with a mean peaked at 686 keV.
The difference between 565 keV and 686 keV can be partly explained by the large binning used for these energies because
of the logarithmic progression of the bins in order to cover a wide energy spectrum of nine decades. Also the low energy resolution
originated in the response matrix can be invoked to explain this discrepancy and finally too a systematic bias of about 40
keV from the energy calibration of this Van de Graaf generator.
The shape of this spectrum is obviously peaked even if the FWHM is quite large : 726 keV.
The global statistics errors represents 19. \% of the unfolding spectrum (see figure 2a) while the systematics errors cover
28.8 \% (see figure 2b); even if these two figures are quite large, this kind of spectrum estimation is sufficient enough for radioprotection
purposes. These large values of uncertainties are probably due to the low counting rates compared to the thermal neutron
measurement and also to the bad signal to background ratio for direct and diffused neutrons measures.
Figure 2d illustrates the high speed of the unfolding process: after only 97 iterations, the stability is obtained.

\subsection{Evolution of the minimization criteria $Q_\lambda$ for thermal neutron unfolding}
The progress of the minimization criteria  $Q_\lambda$ for the unfolding of the thermal neutron from Sigma
is illustrated in figure 3. The $Q_\lambda$ {\em vs.} $\lambda$ function is drawn after 1, 2, 3 and finally 78 iterations.
The evolution of this minimizing criteria is very fast.
After the first iteration a wide minimum can be seen, while after the second iteration it is already very faint and extremly closed to 1.
With the evolution of the iteration number the value of this minimum get smaller and smaller and also closer to 1.
The value of  $Q_1$ is always $+\infty$ this explains why the $\lambda$ minimum is very closed to 1 but never equals 1.
So for the last iteration before stability, i=78, we find $1-\lambda \approx 10^{-16}$ .
This kind of unfolding scenario is typical of the cases we have studied, and also the  order of magnitude of the
last value of $1-\lambda \approx 10^{-16}$.

\section{Modification of the interpolation method for the thermal neutron part of the spectrum}
\subsection{Precise study of the thermal part of the unfolded spectrum of neutrons from the Sigma installation}
If we consider into details the unfolded spectrum from the Sigma installation, we can be puzzled by the
anomalous behaviour of this spectrum in the range $10^{-8} -10^{-7}MeV (0.01 -0.1 eV)$. Normally, the curve
should look like a Maxwellian distribution with a temperature of 300K.
\begin{eqnarray}
&{ dN \over dE} \propto E^{1/2}e^{-E \over kT}
&kT=300K=0.025 eV
\end{eqnarray}
Instead of this behaviour, the neutron spectrum distribution seems to rise very strightly at low energies without
 exhibiting a maximum around 25 meV. Three explanations can be argued for this fact. First of all, the low energy
cut of the matrix response and of the unfolded spectrum is put at 10 meV $(10^{-8}eV)$ which
should distort a little bit this low energy region behaviour of thermal spectra. The main reason of this artificial
cut at 10 meV is due to the very bad uncertainties of the response functions of the Bonner Sphere system to
very low energy neutrons produced  by the MCNP monte-carlo program in the range 1meV - 10 meV. So, very large
statistical error bars are obtained through this method at 1 meV (bigger than 20\%).
This bad quality of results produced by MCNP for low energies neutrons is originated by the importance of the kinetic
effect that polyethylen for example can generate through thermal exchanges between the molecules of polyethylene,
whose agitation and binding energy are greater than 22 meV, and the impinging neutrons of very low energy 1 meV.
Thus, these very cold neutrons are heated by the brownian motion of the molecules of the detector and a good deal
 of these neutrons do not penetrate inside the detector but are preferentially bouncing at the first contact with
the surface of the sphere. This explanation is also at the origin of the discontinuities observed on the matrix response
while the diameter of the spheres varies between the bare detector and the first 3 inches sphere for energies lower than
1 eV (see figure 4).
The bouncing neutrons, due to thermal exchange between them and polyethylene explains the reheating and
freezing neutrons between 1 meV and 1 eV. It produces for the spheres of diameter ranging between 1.3 to 2.5
inches at some energies an enhancement of the detector's efficiency and a reduction at other energies
(see figure 5).
This last point brings serious consequences while big discontinuities appears at low energies for small spheres
 between naked counter
and the 3 inch sphere: we are not allowed to perform an interpolation  of the measured data for hypothetical
spheres in the range of the bare detector up to the 3 inch sphere.
(Note : diameter of the naked counter =1.29 inch)
This is equivalent to say that the measurement function is not smooth enough for spheres with diameters lower than 3 inches
to authorize a direct interpolation. A solution would be to perform some more measurements with spheres of lower
diameters than 3 inches, but due to these heavy discontinuities, small variations in diameter (less than 1mm) on the
polyethylene thickness  should generate strong variations on the neutron's rates.
\subsection{Rectifying the interpolation}
 We propose an alternative solution to this problem :
in most of the actual cases, the low energy spectrum follows a Maxwellian distribution; the very rare cases where this
assumption is wrong are very few and are well known situations as for example : a specific low energy neutrons filtered beam
produced with a nuclear reactor adapted to solid state physics and cristallography. In other words, guessing a Maxwellian distribution
is a low cost, extremly realistic and highly efficient  assumption for low energies neutrons encoutered in actual radioprotection situations.
So we performed an interpolation of the measurements for the spheres between 2.9 inches up to 12 inches, while
we generated for a pure maxwellian distribution response of the spheres  lower than 1.9 inches down to
the naked counter, matching the data measured with this bare counter and its response to this pure Maxwellian distribution.
For the hypothetical spheres between  1.9 and 2.9 inches, we performed a linear transition to connect these two
curves of reconstructed measurements versus diameters of the spheres.
Unfolding with this method the Sigma spectrum shows a satisfying solution regards to the Maxwellian distribution.
The price to pay is a loss of continuity between the thermal and the epithermal parts of the unfolded spectrum. Anyway,
the epithermal part is rather small compared to the  pure thermal part and secondly the incidence of this epithermal part
is rather small on the total dose flux measurement, because it deals with few neutrons ranging between 1 and 100 eV.
The interesting point of this method is that it doesn't affect at all the unfoldings of greater energies neutrons spectra,
for example the 565 keV neutrons monoenergetic spectrum is absolutely unchanged!
\subsection{Revisiting the Sigma thermal neutron unfolded spectrum in the light of the interpolation rectification}
 Using the rectification of the interpolation method with a Maxwellian distribution forcing for the thermal part of the spectra,
onto a real data set from the Sigma thermal neutron facility, Am-Be sources + graphite moderator (see figure 6c).
 We obtain a very interesting neutron unfolded spectrum which exhibits a clear Maxwellian distribution for
thermal neutrons of energy below $10^{-7}eV (100meV)$, a high energy part of the spectrum for neutrons
ranging from 1MeV up to 10 MeV corresponding to direct neutrons produced in the Am-Be source (see figure 6a).
Between 10 and 100 eV $(10^{-6}-10^{-4}MeV)$, we can find an enhanced epithermal band, that seems to be disconnected
from the Maxwellian thermal distribution. Here is the price to pay for this linear combination of two method
of data interpolation (Spline+Maxwellian forcing) for the small diameters sphere countings.Anyway, this small epithermal
band is not of a very importance for the dose flux calculation.
Also, the global statistics errors remains very low: $0.24\%$ , compared to $0.11\%$ previously(see figure 6a).
While the global systematics errors is slightly increased from $3.6\%$ to $5.15\%$ (see figure 6b).
We can see that the final distribution is obtained after 85 iterations with a minimizing parameter  $1-\lambda $
less than $10^{-11}$. The modification of the interpolation gives more realistic results when the original approach
shows a small deficience due to discontinuities while interpolated. Fortunately we have a very good knowledge,
thanks to the thermodynamic of the standard thermal distribution of low energies neutrons.

\section{Conclusion}

The development of this unfolding iterative Tykhonov approach was leaded by three important requirements :
a high efficiency and fast converging method, a better and faster convergency than the least
square method or than the classical Tykhonov method, but with the stability of the solution toward
fluctuations of the measurements produced by the Tykhonov approach and also, a complete unsensitivity of the solution with respect to the initial guess.
The iterated Tykhonov unfolding approach is modified to ensure a good convergence with a parameter
running between 0 and 1.
An accelerating algorithm is introduced between two classical Tykhonov iterations, this algorithm looks like a gradient
method with a least square term.
The calculation of statistics and systematics uncertainties is studied in details.
The efficiency of this approach is presented in two experimental neutron spectrum unfoldings.
This method is fully dedicated to Bonner Sphere system unfolding.
Due to its very wide energy coverage, the Bonner sphere system is a very usefull device for radioprotection evaluation,
but it cannot be used as a high resolution neutron spectrum instrument for metrology, even with an excellent
unfolding method!
One of the very big advantages of the method here illustrated is that no a priori information is introduced during
the unfolding process leading to coherent results with the expected spectra. This independance toward the initial guess
is extremely important for radioprotection measurement when a-priori no information on the neutron spectrum is available;
not observing this point should lead to very prejudiciable wrong evaluations of dose flux calculation which is not allowed
for radioprotection expertise approach. Nevertheless a very slight stretch to this law is applied
in such a way that the interpolated counting rates between naked counter and the first 3 inches sphere is assumed
to correspond to a pure Maxwellian distribution for the thermal neutrons part. This rectification is done to prevent
from the high discontinuities of the matrix response for low diameters and low energies. This type of unfolding
gives realistic results for thermal neutrons without modifying the high energies neutron spectra.

\ack{We must thank for all their help to build the Bonner Sphere, perform measurements, calibrations and matrix response calculation
all the personal from the "Laboratoire de Recherche en Dosim\`etrie Externe" of the IRSN and more specially Dominique Cutarella,
Christian Iti\'e and Emeric Pierre. We must also thank the PTB group of Germany and more specially M. Alevra who helped
us to build this system C Bonner Sphere system.
Most of this work was done at IRSN(Institut de Radioprotection et de S\^uret\'e Nucl\'eaire (IRSN)
 Fontenay aux Roses (France) since 1997 up to 2002 at the Laboratory of External Dosemetry.}

\begin{figure}
\begin{center}
\epsfxsize=15.cm
\epsfbox{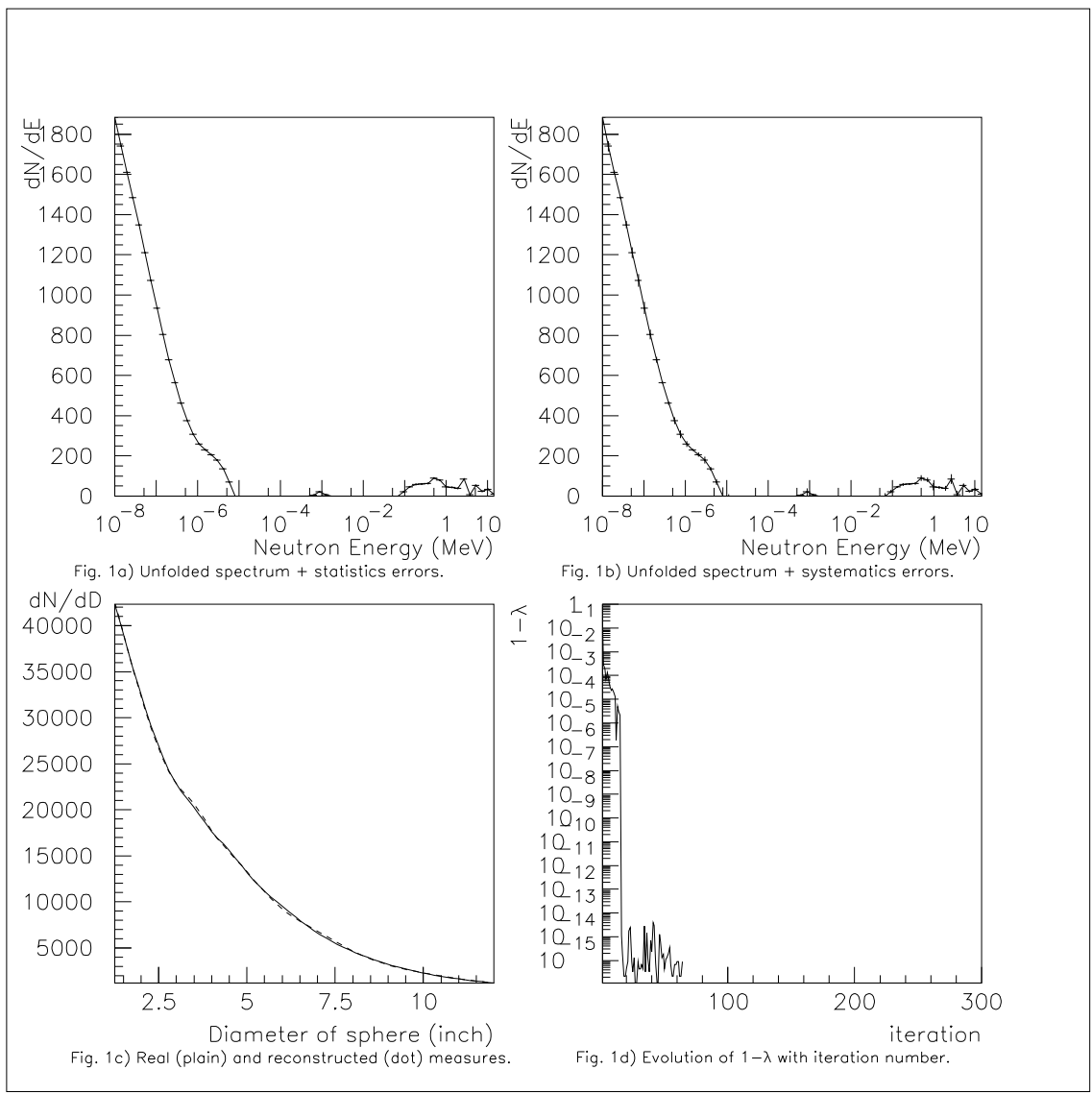}
\end{center}
\caption{\label{fig :sigma spectrum unfolding}
Sigma spectrum unfolding}
\end{figure}

\begin{figure}
\begin{center}
\epsfxsize=15.cm
\epsfbox{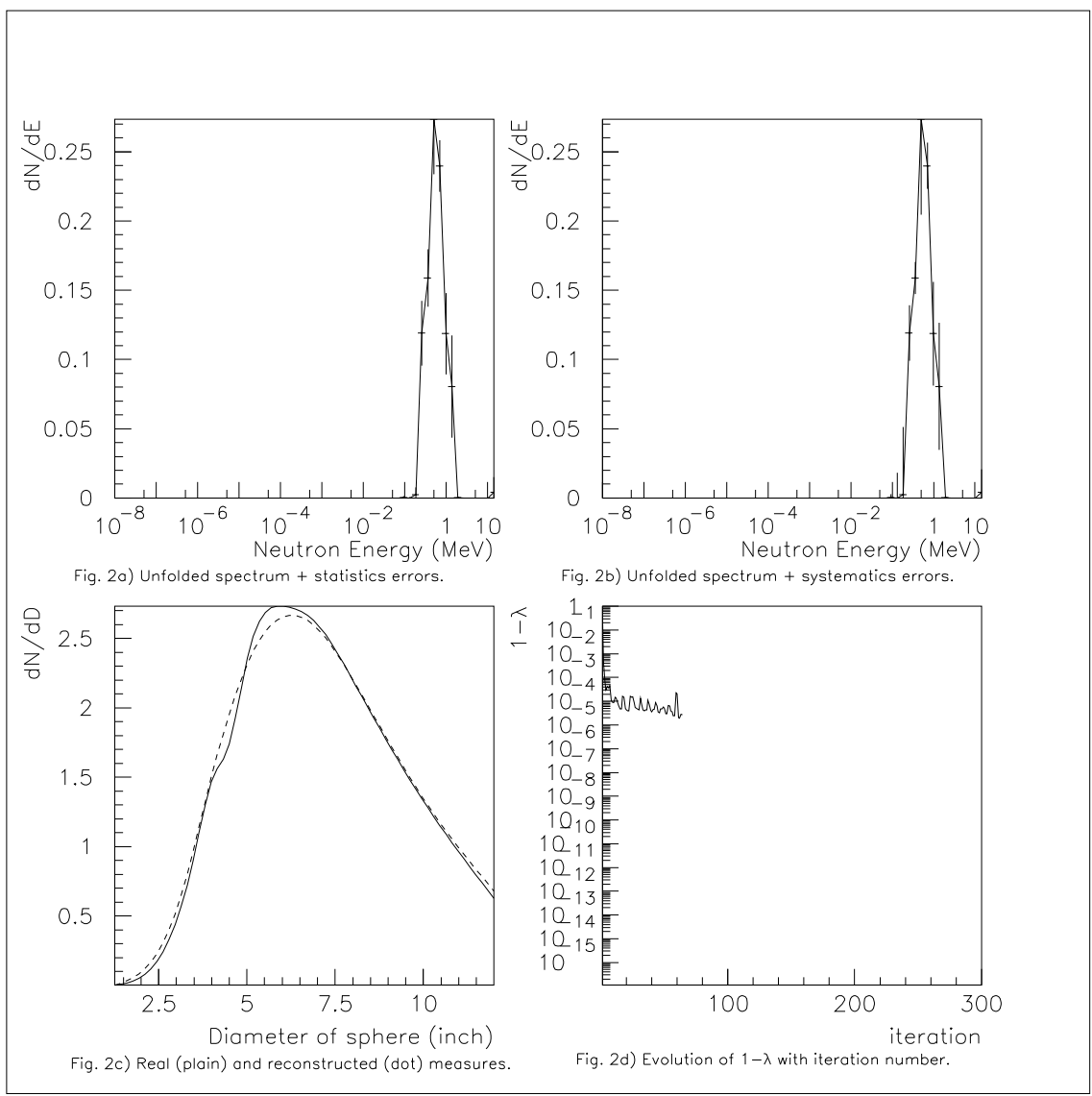}
\end{center}
\caption{\label{fig : 565 keV spectrum unfolding}
565 keV spectrum unfolding}
\end{figure}

\begin{figure}
\begin{center}
\epsfxsize=15.cm
\epsfbox{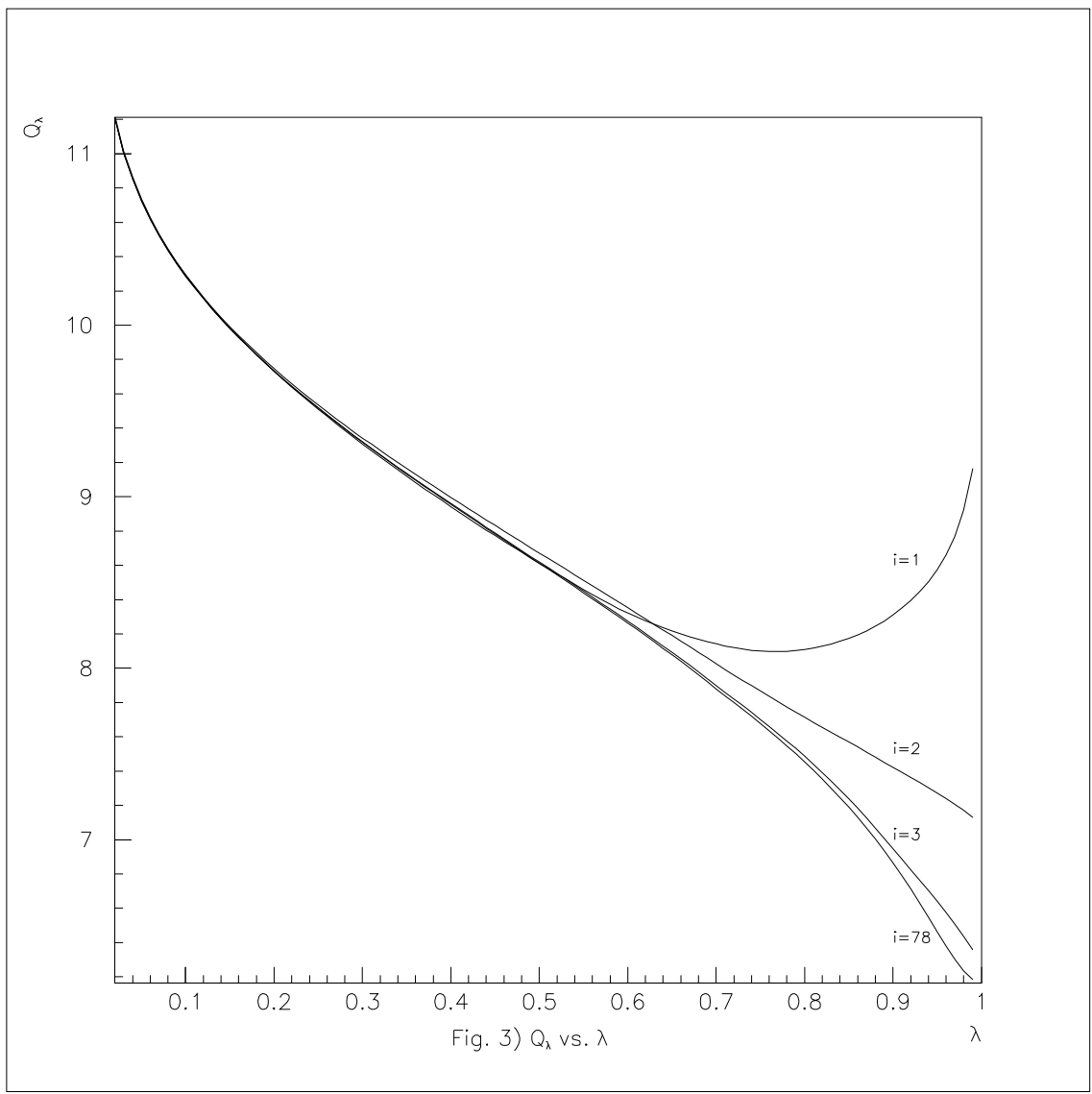}
\end{center}
\caption{\label{fig : Evolution of qlambda}
Evolution of $Q_\lambda $ {\em vs.} $ \lambda$ with iteration number}
\end{figure}

\begin{figure}
\begin{center}
\epsfxsize=15.cm
\epsfbox{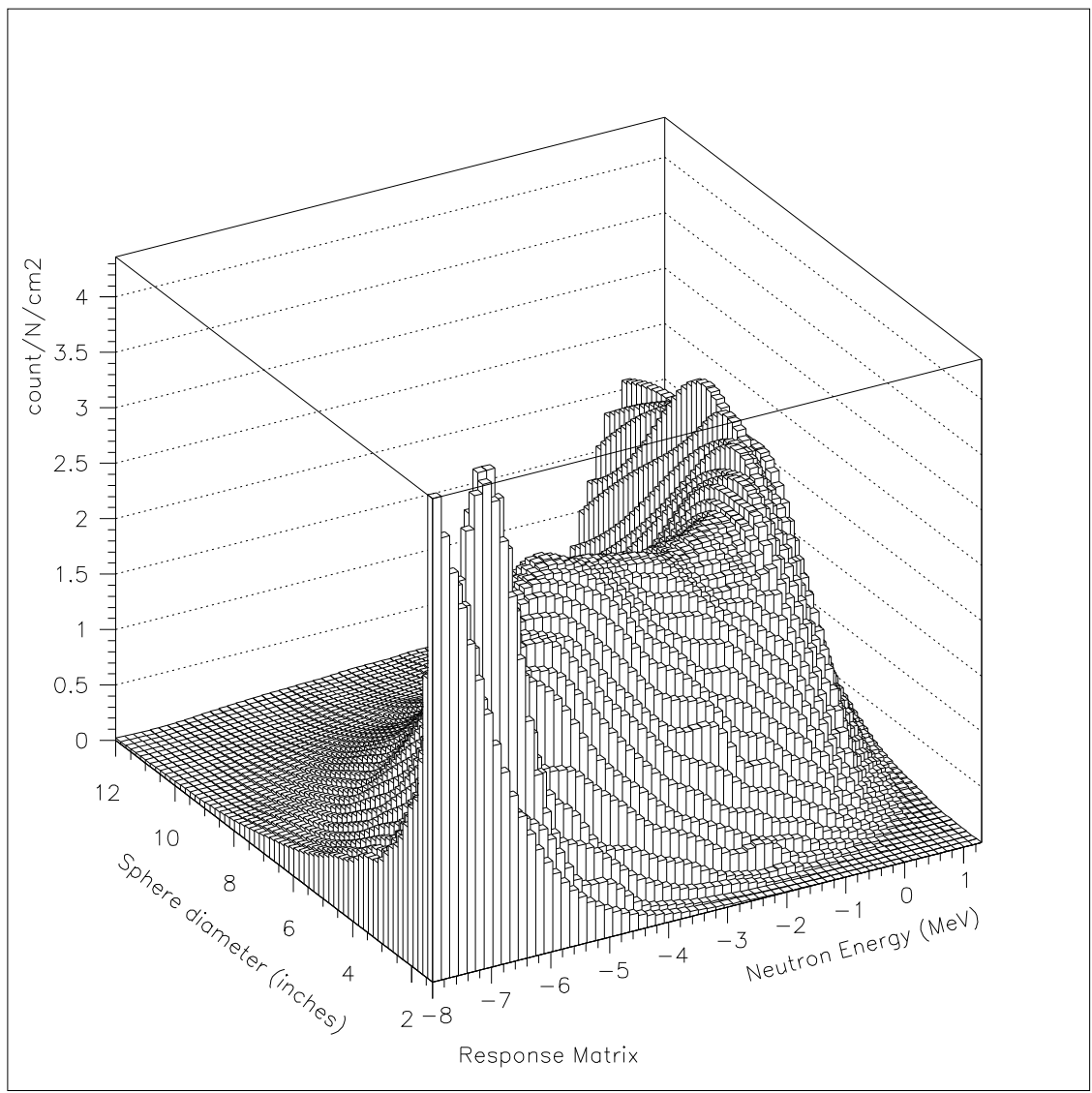}
\end{center}
\caption{\label{fig : Matrix Response of Bonner spheres}
Matrix Response of Bonner spheres versus $log_{10}(NeutronEnergy/Mev)$ versus Sphere diameter (inch)}
\end{figure}

\begin{figure}
\begin{center}
\epsfxsize=15.cm
\epsfbox{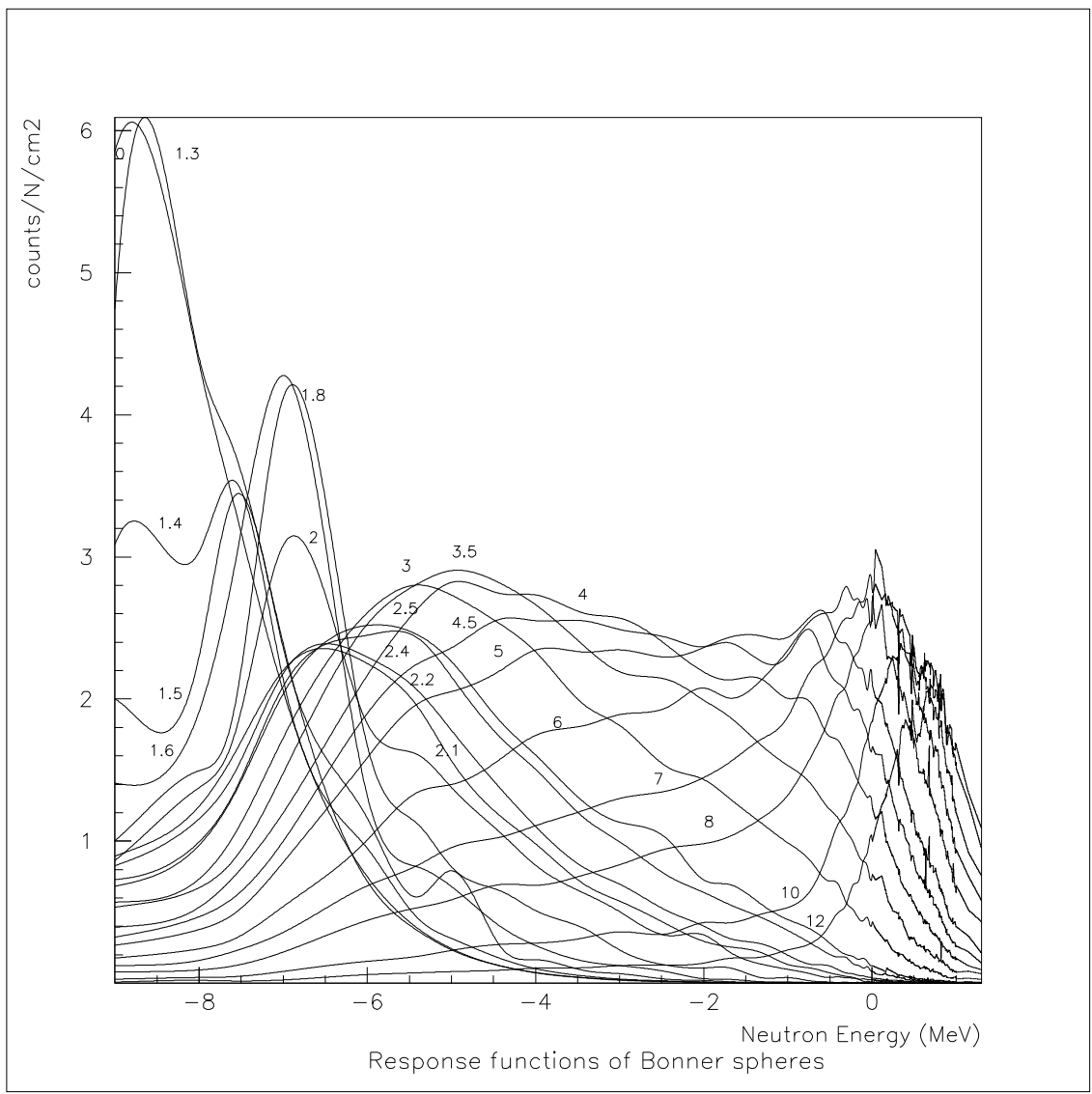}
\end{center}
\caption{\label{fig : Response functions of Bonner spheres}
Response functions of each Bonner spheres versus $log_{10}(NeutronEnergy/Mev)$}
\end{figure}

\begin{figure}
\begin{center}
\epsfxsize=15.cm
\epsfbox{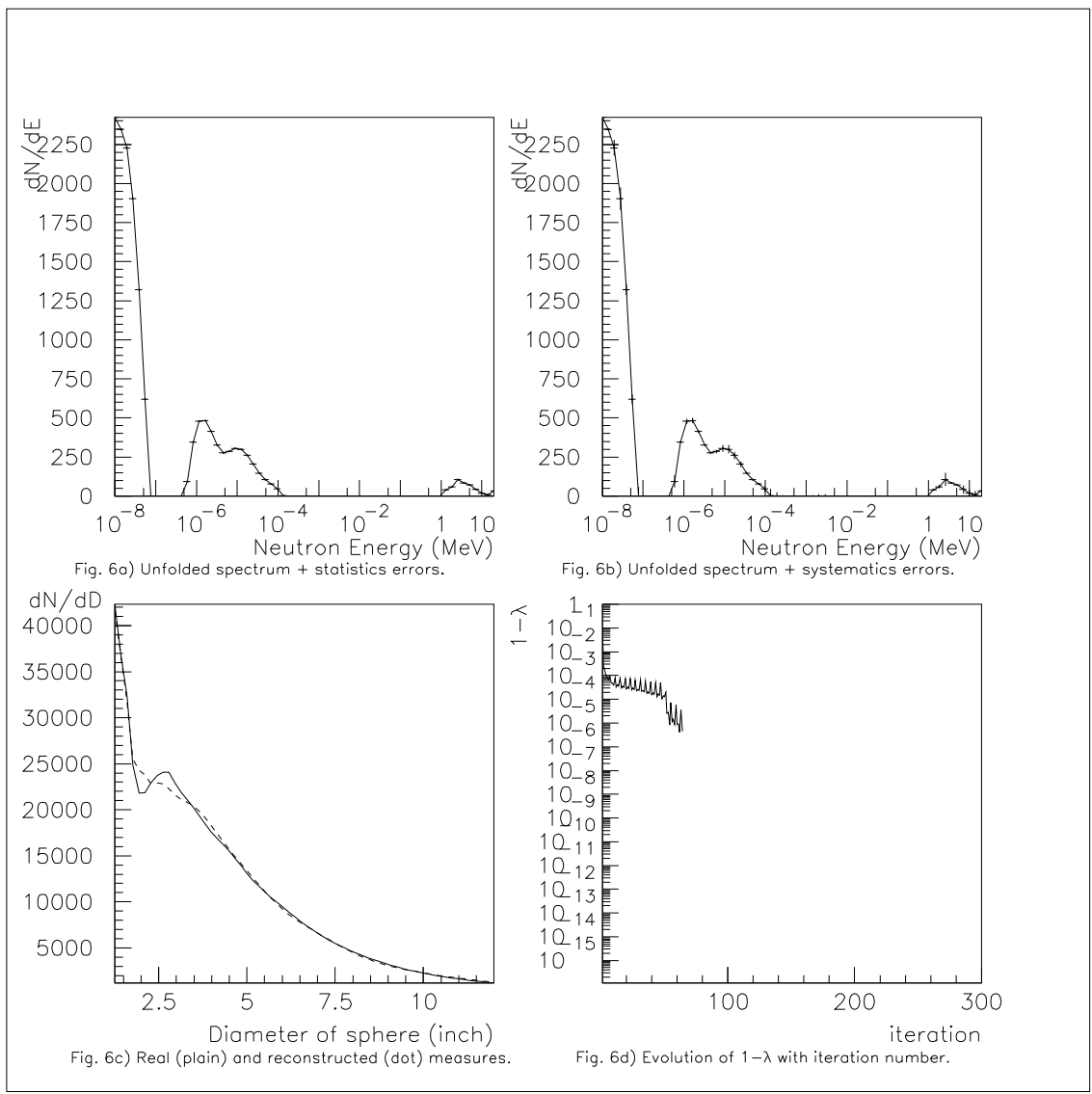}
\end{center}
\caption{\label{fig : Sigma spectrum unfolding with modified interpolation
: linear combination of Spline + Maxwellian distribution forcing in the thermal part}
Sigma spectrum unfolding with modified interpolation
: linear combination of Spline + Maxwellian distribution forcing in the thermal part}
\end{figure}

\end{document}